\documentclass[twocolumn]{aa}
\usepackage{graphicx}
\usepackage{epsf}
\usepackage{pdflscape}
\usepackage[figuresright]{rotating}
\usepackage{txfonts}
\begin{document}
   \title{Optical spectroscopy of close companions to nearby \\
          Herbig Ae/Be and T Tauri stars
   \thanks{Based on observations collected at the European Southern
Observatory, Chile (program ID 075.C-0395(A))}}
\titlerunning{Optical spectroscopy of close companions to PMS stars}

   \author{A. Carmona
          \inst{1,2}
          \and
          M.E van den Ancker \inst{2}
	  \and
          Th. Henning \inst{1}
          }

   \offprints{A. Carmona\\
              \email{carmona@mpia.de}}
             
   \institute{
              Max Planck Institute for Astronomy, K\"onigstuhl 17, 69117 Heidelberg, Germany
           \and  
              European Southern Observatory, 
              Karl Schwarzschild Strasse 2 , 85748 Garching bei M\"unchen, Germany    
             }

   \date{accepted by A\&A 18/12/2006 doc\#AA/2006/5509}

   \abstract{
     We present VLT-FORS2 optical (5700 - 9400 \AA) spectroscopy of
     close ($r$ $<$ 1.5'') companions to three nearby (d $<$ 200 pc) Herbig Ae/Be stars
     (HD 144432, HD 150193, KK Oph) and one T Tauri star (S CrA).
     We report the detection of Li I (6707 \AA) in absorption and emission lines
     (H$_{\alpha}$, Ca II triplet) in the spectra of the companions.
     Our observations strongly suggest that the companions are physically associated pre-main-sequence stars.
     The spectral type derived for the companions is  
     K5Ve for HD 144432 B, F9Ve for HD 150193 B, and G6Ve for KK Oph B.
     S CrA A and B were observed simultaneously. 
     The spatially resolved spectra indicate that S CrA A (primary, north) is a G star 
     and that S CrA B (secondary, south) is a K star.      
     Using photometry from the literature and 
     estimations of the $R$ and $I$ magnitude derived from the spectra, 
     we localized primaries and companions in the HR diagram, derived their masses 
     and assuming coevality constrained the age of the systems.
     KK Oph B (7 Myr) and S CrA B (2 Myr) are actively accreting T Tauri stars and are very likely surrounded by disks.
     HD 150193 B (10 Myr) and HD 144432 B (8 Myr) are weak-line T Tauri stars.     
     Three of the four systems studied (HD 144432, HD 150193, KK Oph) have ages $>$ 7 Myr.
     These systems retained their disks for a longer time than typical of a young star.
     Our results suggest that binarity may be a key issue in understanding the lifetime of disks.

   }

\maketitle

%
\begin{table*} %
\caption {Log of the observations. 
The first column lists the name of the primary.
The second and third list the separation $r$ in arcsec and 
the position angle $P.A.$ in degrees of the companion with respect to the primary.
$T_{exp}$ refers to the integration time in each grism.
H$\alpha$/cont is the H$\alpha$ peak/continuum ratio in the primary.
Max. contamination refers to the maximum level of contamination in the continuum in the companion's spectra
(see \S 3.1). 
References: [PE04] P\'erez et al. 2004, [PI97] Pirzkal et al. 1997,
[BA85] Baier et al. 1985.  
 }
\label{table:1}      
\centering                          
\begin{tabular}{lllllllccllllll}  
\hline\hline                 
Primary      & $r$        &  P.A         & Ref & seeing & airmass & $T_{exp}$ & H$\alpha$/cont & Max. contamination\\
             & [arcsec] &  [$^{\circ}$] &     & [``] &         & [s] &   &[\%]      \\
\hline %
HD 144432 & 1.4 & 4 $^a$ & PE04 & 0.7 & 1.1 & $3\times60$   & 3.4  & 7\\
HD 150193 & 1.1 & 236 & PI97 & 0.6 & 1.4 & $3\times11$   & 2.2  & 8\\
KK Oph    & 1.5 & 257 & PI97 & 0.9 & 1.0 & $3\times110$  & 7.3  & 2\\
S CrA     & 1.4 & 157 & BA85 & 0.6 & 1.1 & $1\times110$  & 19.3 & 2\\
\hline %
\end{tabular} %
\begin{flushleft}
$^a$ PA re-determined from a new analysis of the IRTF image shown in P\'erez et al. (2004).
\end{flushleft}
\end{table*} %
%
%
\begin{figure*}
\centering
\includegraphics[angle=0,width=\textwidth]{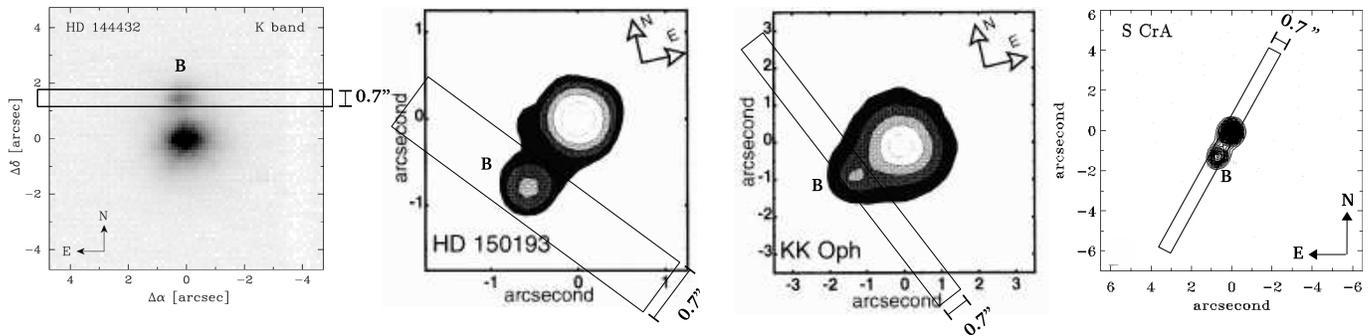}
\caption{Slit positions and the targets observed. 
The slits have the sizes scaled to the image obtained from the literature. References: HD 144432 [PE04], HD 150193 \& KK Oph [PI97], 
S CrA (Reipurth \& Zinnecker 1993).}
\label{Slit configuration}
\end{figure*} 
\section{Introduction}
Planets form in disks of gas and dust around stars in their pre-main-sequence phase.
Given that our Sun is a single and isolated star,
investigations have been biased towards the study 
of planet formation around similar objects.
In contrast, 
statistical studies reveal that
more than 50\% of the solar-type main-sequence stars in the solar neighborhood
are members of a binary or a multiple system 
(e.g., Duquennoy \& Mayor 1991; for a recent review see Bonnell et al. 2006)
and that the multiplicity fraction is significantly higher in pre-main-sequence stars 
(e.g., Reipurth \& Zinnecker 1993; Pirzkal et al. 1997; for a recent review see Duch\^{e}me et al. 2006).
In addition, 
investigations of star-forming regions and young stellar clusters 
show that most stars form in groups, 
and searches for extrasolar planets demonstrate that giant gaseous planets 
do exist in binary and multiple systems 
(e.g., Konacki 2005; for a recent review see Desidera \& Barbieri 2006).

How planetary systems form and evolve in binary and multiple systems
is an important question that is starting to be addressed theoretically 
(e.g., Kley 2000, Nelson et al. 2003, Moriwaki \& Nakawa 2004, Boss 2006, Quintana \& Lissauer 2006). 
From the observational perspective,
besides continuing the search for planets in multiple systems,
one logical step after the numerous imaging surveys for multiplicity in pre-main sequence stars
(e.g., Chelli et al. 1988; Reipurth \& Zinnecker 1993; 
Ghez et al. 1993, 1997; Li et al. 1994; Leinert et al. 1993, 1997; 
Richichi et al. 1994; 
Simon et al. 1995; Pirzkal et al. 1997; Bouvier \& Corporon 2001; Koresko et al. 2002;  Kouwenhoven et al. 2005; 
Ratzka et al. 2005) is to better constrain the characteristics of the companions of nearby sources with resolved disks.
Binary systems with disks may be the progenitors of the binary systems with planets.
The study of these systems when they are very young and still posses a disk will allow 
us to unveil around which component planets are forming, or whether there can be such a thing as circumbinary planets.
To attain such goals, the first steps are: to obtain further evidence that the companion is not a background object, 
to constrain the spectral type (mass) of the companion(s),
and to derive an independent estimate of the system age.   

In this paper we describe an effort to characterize close ($r$ $<$ 1.5'' in projection) visual companions to four nearby ($d$ $<$ 200 pc)
pre-main-sequence stars surrounded by protoplanetary disks: 
the Herbig Ae/Be stars (HD 144432, HD 150193, KK Oph) and the T Tauri star S CrA.
We seek to determine whether these companions are physically associated to the primary,
or whether they are background sources. 
We derive the spectral type of the companions, 
place primaries and companions in the Herzsprung-Russell diagram (HRD)
and show that in all cases these companions are physically associated with the primary.
The paper is organized as follows: In \S2 the observations and the data reduction are described, 
in \S3 the methods of spectra classification and luminosity determination are presented,  
and in \S4 our results are discussed.
\section{Observations \& data reduction}
\subsection{Observations}
During the night of 17 May 2005 we observed the companions to
HD 144432, HD 150193, KK Oph and S CrA  
with the FORS2\footnote{http://www.eso.org/instruments/fors2}
optical spectrograph mounted on the VLT at Paranal Observatory in Chile.

The observations proceeded as follows: 
(1) The telescope was pointed to the primary
and the instrument was rotated 90$^{\circ}$ with 
respect to the line joining primary and secondary. 
A first acquisition image was taken. 
(2) A blind offset in the direction of the secondary was performed and an image 
through the slit was taken.
The blind-offsetting values were calculated based on the separation and position 
angle published in the literature (see Table 1).
(3) Small adjustments (1-2 pixel) were performed in the pointing for ensuring that 
the companion was well centered on the slit.
(4) Two science exposures were taken, one using the grism 1200R (wavelength 
range 5740-7230 \AA,~R$\approx$2140) 
and another one with the grism 1028z (wavelength range 7660-9400 \AA,~R$\approx$2560).
In each grism exposure three integrations were executed, 
each in a different position on the CCD for avoiding errors due to 
the influence of bad pixels. 

The slit was oriented perpendicular to the line connecting the primary and the secondary 
(see Fig. 1).
Given the difference in brightness,
a perpendicular slit orientation
minimizes the amount of light of the primary that falls on the
CCD - thereby greatly alleviating problems of dynamic range,
saturation and electronic cross-talk.

In the case of S CrA (A\&B), however, the slit was oriented in the direction of the line connecting the 
primary and secondary. 
The small difference in the brightness of the system components (1 mag) 
allowed us to obtain  the spectra of primary (S CrA A) and secondary (S CrA B) simultaneously.
 
Integration times were calculated as a function of the estimated $V$ magnitude of the 
companions.
The slit width was selected as a function of 
(a) the astrometric accuracy of the position of the companion, 
(b) the separation of the sources and (c) the seeing conditions.  
This led to the choice of an 0.7'' wide slit width.

\subsection{Data reduction}
In Table 1 we present a summary of the observations.
To calibrate the data, at the end of the night we observed
the spectro-photometric standard star HD 156026 using the same slit (0.7'')
and the two grims (1200R, 1028z). 
Lamp flatfields and lamp Ar-Xe arcs with the 0.7'' slit for each of the two grims 
were also taken at the end of the night.

Raw data were reduced using standard techniques involving MIDAS, IRAF and IDL.
Multiple exposures for one science target were reduced and extracted independently.
The bias was determined in an homogeneous region in a dark exposure of zero seconds.
Each flatfield frame was normalized after bias subtraction. 
A master flatfield was created by taking the median of all the flatfield frames. 
Each raw science frame was corrected for differences in the pixel gain 
by dividing by the master flatfield after the subtraction of the bias.

Each flatfield-corrected science frame was tilt corrected and simultaneously wavelength calibrated 
using the arc-lamp exposures.
Standard IRAF routines were used for the tilt correction as described in the 
``IRAF User's Guide to Reducing Slit Spectra''
by Massey et al. (1992).
The sky lines and cosmic rays were subtracted from the 
tilt corrected frames before the spectral extraction.
This procedure was particularly necessary for the case of S CrA (A\&B),
because the two sources overlapped in the spatial direction and 
the  extraction algorithm required the selection of very small (2-4 pixels) background windows.

The extraction of the spectra of HD 150193 B, KK Oph B, HD 144432 B 
was performed in a straightforward manner, 
using the standard optimal-extraction algorithm EXTRACT/LONG implemented in MIDAS. 
In this method, two background windows  and one extraction window 
are selected in the spatial direction of the spectra.
The algorithm calculates the background correction and performs the extraction, assigning different 
weights to each line where the spectrum is extracted.
The final spectrum is the weighted sum of each contribution in the spatial direction.
Identical results were obtained with the routine ``apall'' of IRAF.
Finally, the individual wavelength-calibrated one-dimensional spectra from each 
exposure were added to obtain a final spectrum. 
Features present in only one of the exposures were considered reduction 
artifacts (e.g unfiltered cosmic rays) and were removed from the final combined spectrum.
%
\begin{figure}
\centering
\includegraphics[angle=0,width=0.5\textwidth]{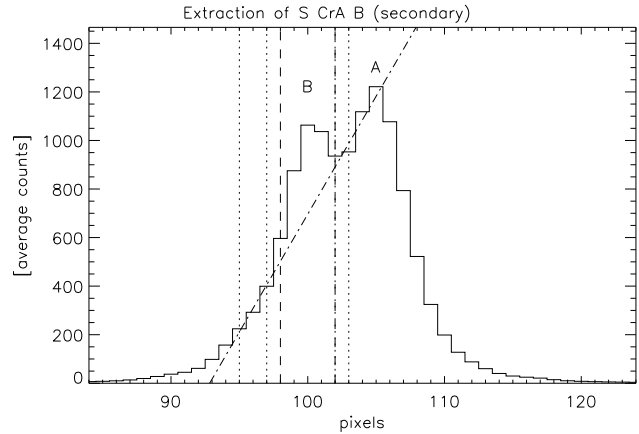}
\caption{Example of the extraction method in 
the case of S CrA A and S CrA B.
Here we show a cut in the spatial direction of the spectra at an arbitrary wavelength.
The PSF of the companion (left)  overlaps with the PSF of the primary. 
In the figure, the vertical dashed lines represent the limits of the target 
extraction window (in this case the companion S CrA B)
and the vertical dotted lines show the limits of the background windows.
In the graph, the right limit of the extraction window and the left limit of the right background window overlap. 
The dot-dashed diagonal is the background fitted
(see \S 2.1 for discussion). 
The first peak from the left is S CrA B (companion), the second is S CrA A (primary).
The pixel scale is 1 pixel = 0.252''.
}
\label{Extraction}
\end{figure}
 
In the case of the more complex reduction of S CrA (A\&B) we developed an 
interactive IDL algorithm for extracting the spectra. 
The extraction procedure used the following steps: 
(a) the 2D frame is averaged in the wavelength direction and from the resulting spatial profile
two background windows and one extraction window are chosen, 
(b) at each wavelength, the background is determined by a linear fit to the pixels lying in the background windows
(chromatic background correction),
(c) at each wavelength, 
the respective fitted background is subtracted to each pixel in the spatial direction,  
and (d) at each wavelength, all pixels in the extraction window are summed. 
An interactive routine was required because the best extraction demanded a careful 
tuning of the background and extraction windows.
The lengths of the background and extraction intervals were changed pixel by pixel manually 
until the best signal to noise in the spectrum was obtained ($\sim$ 20). 
In Figure 2 we present an example of the extraction of S CrA B (companion).
This procedure offered the best individual spectra of the sources.
Nevertheless, we note that the flux fidelity is not conserved. 

This method is optimized to obtain an optimal S/N ratio 
while preserving a low level of contamination
from a neighboring source when the PSF of two adjacent sources overlap.
We tested the Gaussian deconvolution method for extracting the spectra. 
However, the results obtained (in terms of discerning the individual spectra), 
were not as good as the results obtained with straight-line fitting to the continuum.
In particular the straight-line method was able to show that mostly all Paschen emission come from the 
primary S CrA A (see Fig 3).
We tested a third method consisting of mirroring the right wing of the PSF of the primary as an estimation
of the background of the companion at each wavelength, and vice-versa. 
The spectra obtained for the primary and secondary were almost identical as the spectra obtained with 
the straight-line method. 
However,
the straight-line method offered a better S/N ratio in the companion's spectra in the wavelength region 8800 - 9400 \AA. 

\section{Results}
\subsection{Spectral classification}
In Figure 3 we show the extracted spectra of the companions.
In all the companions we detected H$_{\alpha}$ (6563 \AA) in emission and Li I at 6708 \AA~ in absorption.
Ca II (8498, 8542, 8662 \AA) in emission is detected in all the companions except for HD 144432 B.
Details of important spectral features are illustrated in  Figure 4.
To determine the spectral type of the sources, we employed two complementary methods.
First we applied the quantitative method of Hernandez et al. (2004).
Then we compared the spectra with STELIB, 
a library of standard star spectra obtained with a similar spectral resolution (R$\approx$2000; Le Borgne et al. 2003)
\footnote{www.ast.obs-mip.fr/users/leborgne/stelib/index.html}.

The Hernandez et al. (2004) method is based on the measurement of the equivalent width of selected spectral features 
(photospheric in nature and less affected by activity-induced line emission)
and assignment of a spectral type (or range of spectral types) to each spectral index measured.
The combination of several spectral features narrows down the possible 
spectral type(s) to a range consistent with the multiple diagnostics.
The method offers the advantage that when only very few lines are detected (and placing meaningful
upper limits on the presence of others) it is possible to extract an accurate spectral type.
This method is mainly useful for early type stars. 

Calibration curves relating spectral types with equivalent widths of spectral features
were computed by Hernandez et al. (2004). 
In the range covered by our observations 
the spectral features employed for spectral classification were :
He I (5876 \AA), Na I (5890 \AA), Mn I (6015 \AA), Ca I (6162 \AA), 
TiO (6185 \AA), CaH$_2$ (6385 \AA), He I (6678 \AA), TiO$_3$ (6720 \AA),
CaH$_3$ (6830 \AA), CaH (6975 \AA) and He I (7066 \AA).
The equivalent width measurement is presented in Table 2.

{\it Effect of the veiling.} The companions are found to be emission-line T Tauri stars. 
Active T Tauri stars are affected by veiling, a smooth ``continuum'' excess arising from 
the hot (T$\approx$8000-10000 K) accretion spots.
This excess decreases the equivalent width of all absorption features,
potentially affecting the spectral typing using methods relying on quantitative 
measurements of the line strength.
This smooth continuum can contribute significantly to the total flux 
(up to 50-70\% in the red part of the spectrum), 
implying comparable reductions in the equivalent widths (Hartigan \& Kenyon 2003).
For the stars,
for which we have detected many lines such as
HD 144432 B, HD 150193 B, KK Oph B we can exclude the possibility of strong veiling.
Large amounts of veiling would mean that different lines would yield to 
mutually exclusive spectral types.
Since this effect is not observed in the three objects the veiling should be small ($<$ 10\%).         
The case of S CrA A\&B is more complex since fewer lines are detected and the veiling
effect limits the derivation of good upper limits of line strengths.
However, non-detections combined with the detection of few photospheric absorption lines 
give a spectral interval in which the line can reach the measured strength regardless of the amount of veiling. 
Although error bars are larger, 
the determination of a range of spectral types is still possible.

The comparison with STELIB is a good additional method which is more accurate than the Hernandez et al.
method for later spectral types (K and later), in which the quantitative evaluation of - often blended - 
absorption bands becomes difficult.
For the direct comparison between our spectra and the STELIB spectra, 
the spectra were normalized by a second-order polynomial fit to the continuum.
To perform the spectral classification, the depth of the absorption features 
and the overall shape of the spectra were compared to the templates.

In Table 2 the spectral classification results are summarized.
In Figure 3, the extracted spectrum for each companion is plotted together with
comparison spectra of sources in the lower and upper spectral range of classification.
The spectral type range deduced determines the range of effective temperatures. 
The error on $T_{eff}$ is  given by the range of possible spectral types (see Table 4).
\begin{figure*}
\centering
\includegraphics[angle=180,width=0.95\textwidth]{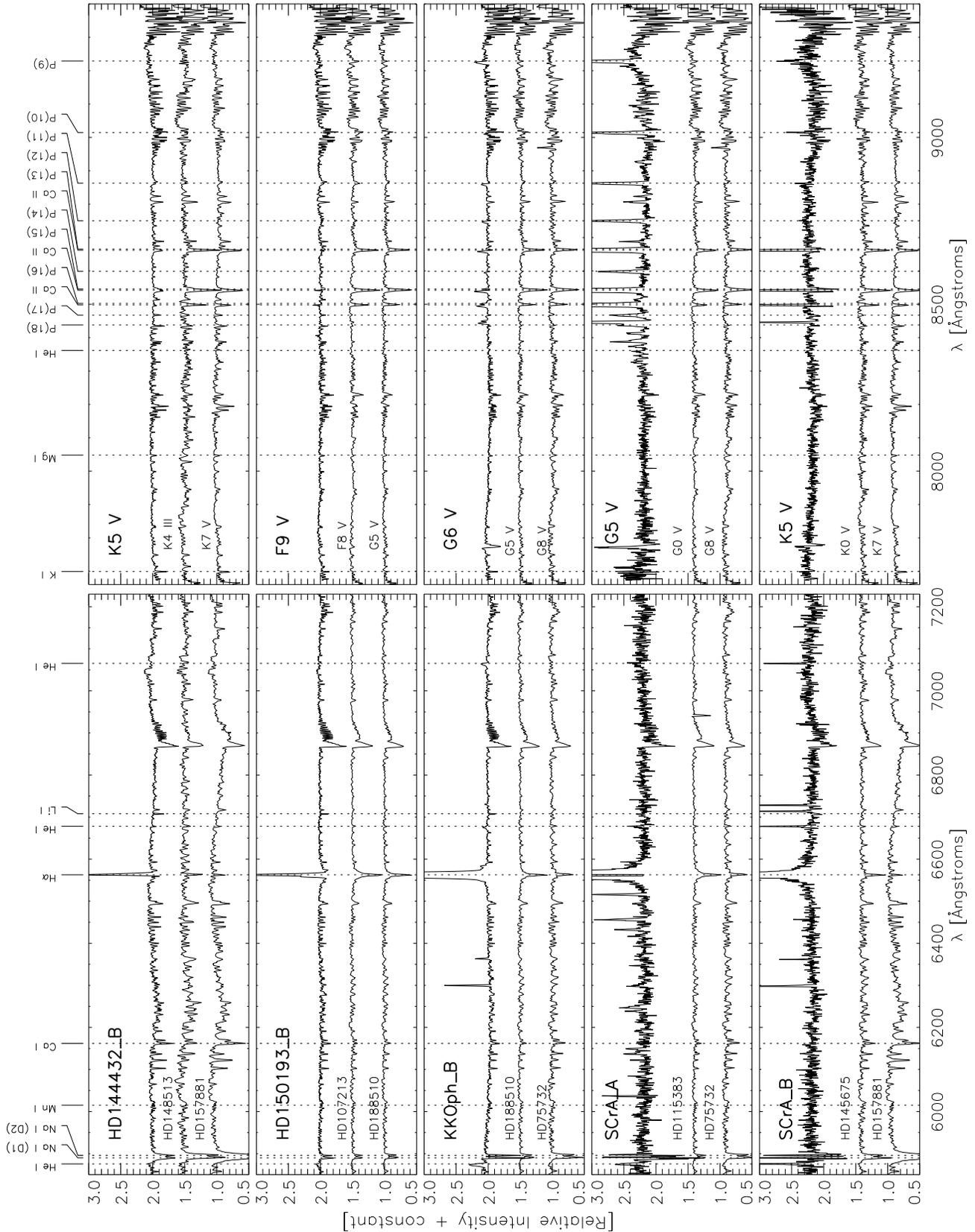}
      \caption{Spectra obtained for the companions. 
               The spectra in the left column were taken with the grism 1200R, 
               the spectra in the right column with the 1028z grism.
                            In each panel, 
               we also show the spectra of standard stars from the STELIB spectral library for the range of consistent spectral types. 
               The derived spectral type of the  companion is indicated in the upper left part of the spectra in the right column.
               The spectra have been continuum-normalized with a second-order polynomial function.
                     }
\end{figure*}
\begin{figure*}
\centering
\includegraphics[angle=0,height=0.65\textheight]{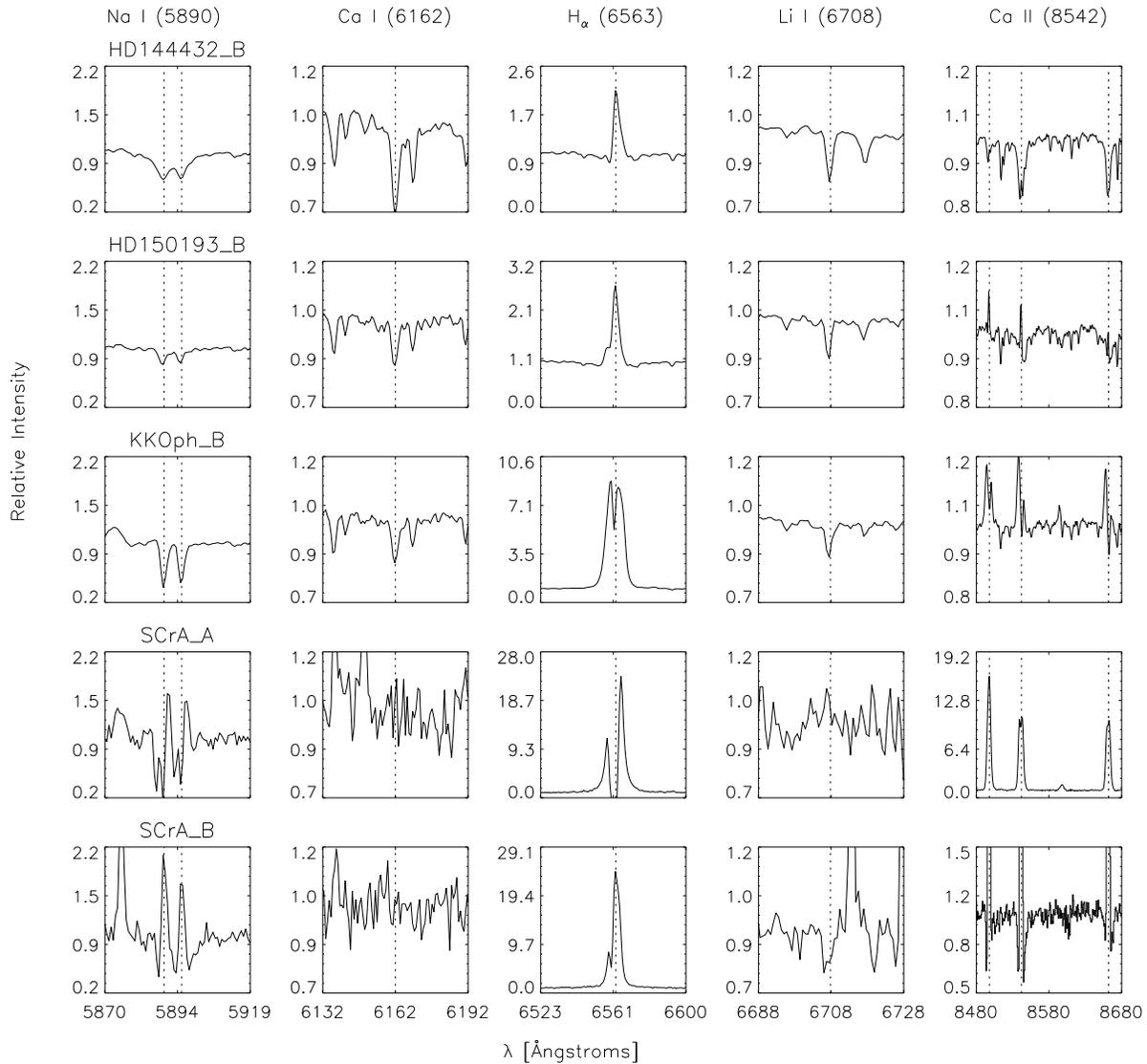}
      \caption{Details of important spectral features (Na I D1 \& D2, Ca I, H$_\alpha$, Li I, Ca II) in the companion's spectra.
              Each row of panels represents the data for one star.
              The columns show each spectral feature.
               }
\end{figure*}
%
\begin{table*}
\caption{Measurements of the equivalent widths (in \AA) of important lines for the spectral classification
of the companions. Negative values indicate that the line is in emission.
Upper limits are given for spectral
lines whose absence allows us to constrain the companion's spectral type.} 
\centering 
\label{table:3}      
\centering                          
\begin{tabular}{l r r r r r r r r r r r  }  
\hline\hline                 
Line  &  $\lambda~ [\AA]$  &  HD 144432 B  &  HD 150193 B  &  KK Oph B  &  S CrA A  &  S CrA B  &  \\
\hline 
He I          &5876 &-0.5 $\pm$ 0.1 & -0.3 $\pm$ 0.1 & -0.9 $\pm$ 0.1 & -1.3 $\pm$ 0.4 & -4.8 $\pm$ 0.4 &  \\
Na I D1       &5890 & 1.3 $\pm$ 0.1 &  0.6 $\pm$ 0.1 &  1.5 $\pm$ 0.1 &  0.5 $\pm$ 0.4 & -0.8 $\pm$ 0.4 &  \\
Na I D2       &5896 & 1.2 $\pm$ 0.1 &  0.5 $\pm$ 0.1 &  1.3 $\pm$ 0.1 &  0.4 $\pm$ 0.4 & -0.1 $\pm$ 0.4 &  \\
Mn I          &6015 & $<$0.1        &  $<$0.1        &  $<$0.1        &  $<$0.4        &   $<$0.4       &  \\
Ca I          &6162 & 1.9 $\pm$ 0.1 &  1.0 $\pm$ 0.1 &  1.0 $\pm$ 0.1 &  $<$0.4        &   $<$0.4       &  \\
TiO           &6185 & 1.3 $\pm$ 0.1 &  $<$0.1        &  $<$0.1        &  $<$0.4        &   $<$0.4       &  \\
CaH$_2$       &6385 & 0.2 $\pm$ 0.1 &  $<$0.1        &  $<$0.1        &  $<$0.4        &   $<$0.4       &  \\
H$_{\alpha}$  &6563 &-2.6 $\pm$ 0.1 & -5.6 $\pm$ 0.1 & -80  $\pm$ 0.1 & -96 $\pm$ 0.4  & -127 $\pm$ 0.4 &  \\
He I          &6678 & 0.0 $\pm$ 0.1 &  $<$0.1        & -0.3 $\pm$ 0.1 & -0.8 $\pm$ 0.4 & -2.0 $\pm$ 0.4 &  \\
Li I (1)      &6708 & 0.4 $\pm$ 0.1 &  0.3 $\pm$ 0.1 &  0.4 $\pm$ 0.1 &  $<$0.4        &  1.3 $\pm$ 0.4 &  \\
TiO$_3$       &6720 & 0.4 $\pm$ 0.1 &  0.2 $\pm$ 0.1 &  $<$0.1        &  $<$0.4        &   $<$0.4       &  \\
CaH$_3$       &6830 & $<$0.1        &  $<$0.1        &  $<$0.1        &  $<$0.4        &   $<$0.4       &  \\
He I          &7066 & $<$0.1        &  $<$0.1        & -0.1 $\pm$ 0.1 &  $<$0.4        & -1.2 $\pm$ 0.4 &  \\
K I           &7699 & 0.4 $\pm$ 0.1 &  0.2 $\pm$ 0.1 &  0.4 $\pm$ 0.1 &  $<$0.4        &  0.5 $\pm$ 0.4 &  \\
Ca II         &8498 & 0.0 $\pm$ 0.1 & -0.4 $\pm$ 0.1 & -1.4 $\pm$ 0.1 & -54 $\pm$ 0.4  & -10  $\pm$ 0.4 &  \\
Ca II         &8542 & 0.6 $\pm$ 0.1 & -0.2 $\pm$ 0.1 & -1.2 $\pm$ 0.1 & -44 $\pm$ 0.4  & -16  $\pm$ 0.4 &  \\
Ca II         &8662 & 0.6 $\pm$ 0.1 & -0.1 $\pm$ 0.1 & -0.9 $\pm$ 0.1 & -42 $\pm$ 0.4  & -14  $\pm$ 0.4 &  \\
\\
\multicolumn{2}{l}{spectral type  range} & K4Ve - K7Ve & F8Ve - G5Ve &  G5Ve - G8Ve & G0Ve - G9Ve & K0Ve - K7Ve &  \\
\multicolumn{2}{l}{\bf adopted spectral type}  & K5Ve & F9Ve & G6Ve & G5Ve & K5Ve & \\
\hline 
\end{tabular} 
\vfill 
\end{table*}
%

{\it Contamination from the primaries}. 
The contamination of the companion's spectra by the primary was determined in the following way.
In each 2D stellar spectrum, 
a cut in the spatial direction, averaged over 20 pixels in the dispersion direction,
at the wavelength of H$_\alpha$ was extracted, 
and the total flux within the extraction window was measured. 
An analogue procedure was performed for a standard star frame obtained with the same slit (0.7") 
at a similar seeing (0.6"),
taking care of scaling the standard star PSF so that it had the same high as the companion's PSF.
Assuming that the seeing does not change significantly between the science and standard star observation, 
the maximum contamination in H$\alpha$ is given by the excess of flux (in percentage)
that the target's PSF has over the unresolved PSF from the 
standard star.
To determine the maximum contamination in the continuum for each star, we divided the
maximum contamination in H$\alpha$ by the H$\alpha$ peak/continuum ratio of each primary 
(see Table 1)\footnote{In the case of KK Oph the seeing of the standard star observation is 0.3" smaller.
The maximum contamination is therefore smaller than the upper limit deduced}.
The H$\alpha$ peak/continuum ratio was obtained by rebinning the
high-resolution spectra used in Acke et al. 2005 (HD 144432, HD 150193), van den Ancker, 
unpublished (KK Oph, S CrA) to R=2500. 
Our results are presented in Table 1.       
The contamination is always smaller than 10 \%.  
In the following sections, we discuss in detail the spectra obtained for each companion.

\subsubsection{HD 144432 B}
The spectrum of HD 144432 B shows a forest of metallic absorption lines characteristic of K type stars.
Moderately strong Ca I at 6162 \AA~ indicates that the star is later than K2.
The strength of the Ca II lines at 8498, 8542 and 8662 \AA~ indicates that HD 144432 B has
a spectral type  not later than K7.
The presence of TiO bands around 7100 \AA~ indicate that the star is mid-K type.
The presence of K I absorption at 7698 \AA~ indicates that the spectral type of HD 144432 B is later than a K4 star.
The spectral range of HD 144432 B is, therefore, between K4 and K7. 
We adopt a spectral type K5Ve.
H$_\alpha$ is observed in emission (EW$\approx$-2.6 \AA~) and 
Li I at 6708 \AA~ is detected. 
The spectral type range K4 to K7 is in agreement with the spectral type reported by P\'erez et al. (2004).

\subsubsection{HD 150193 B}
The spectrum of HD 150193 is characterized by the presence of a forest of weak metallic lines
between 5000 and 6000 \AA.
Weak Paschen absorption lines present in the spectra indicate that the star is of spectral type F.
The strength of the Ca I line at 6162 \AA~ indicates that the star is of early G type.
H$_\alpha$ is detected in emission  (EW$\approx$-5.6 \AA~).
Weak Ca II emission lines are observed inside weak Paschen lines.
Lithium is detected in absorption (EW$\approx$0.31 \AA~).
Weak Mg I at 8047 \AA~ is observed.
Moderately strong Na I in absorption is detected (EW$\approx$0.31 \AA~).
Given that the star shows weak Paschen lines and metallic lines simultaneously, 
HD 150193 B should be a late F type star.
Spectral types between F8 and G5 are consistent with the observed spectrum.  
Bouvier \& Corporon (2001), 
using high resolution spectro-imaging and SED fitting, 
derived a K4 spectral type for HD 150193 B. 
However, in our spectra HD 150193 B does not show the strong spectral features characteristic of K-type stars 
(i.e strong Na I, Ca I, K I, and Ca II lines).
We thus adopt a spectral type of F9Ve for HD 150193 B.
Our data demonstrate that HD 150193 B is a T Tauri star, in agreement with the suggestion of Bouvier \& Corporon (2001).
 
\subsubsection{KK Oph B}
KK Oph B shows weak  metallic absorption lines.
The strength of the Ca I line at 6162 \AA,
the weak Mg I line at 8047 \AA~ and the absence of Paschen absorption lines indicate that
KK Oph B is a G type star.
Strong H$_\alpha$ in emission is observed (EW$\approx$-80\AA~).
Li I is detected in absorption.
Double peaked Ca II at 8498, 8542 and 8662 \AA~is detected in emission.
The range of spectral types obtained for KK Oph B is between G5 and G8.
We conclude that KK Oph B is a G6Ve star.
Recently, Herbig (2005) derived a K-type spectral class for KK Oph B
based on spectro-photometry assuming an identical extinction for KK Oph A
and KK Oph B.
In \S 3.2 we will demonstrate that this is not be the case:
the extinction for KK Oph B is larger than for KK Oph A.
Herbig (2005) does report the presence of the H$\alpha$ and
H$\beta$ lines in emission in KK Oph B, in agreement with our
observations.

\subsubsection{S CrA A \& B} 
The spatially resolved individual spectra reveal that 
S CrA A and S CrA B are emission line stars.
Primary and secondary show H$_\alpha$ in emission. 
S CrA A (primary) has a H$_\alpha$ double peaked profile, 
S CrA B (companion) has a single peaked profile with a small blue shifted peak (see Fig. 4).
Ca II in emission is present in both components. 
The Ca II emission in S CrA A (primary) is much stronger and broader than in S CrA B (companion).
Strong Paschen {\it emission} (13-19) is observed only in the primary S CrA A.
The Na I line is characterized by components in absorption and emission in both sources (see Fig. 4).
Li I is detected {\it only} in S CrA B, where it is significantly broadened (probably due to rotational broadening).
Li I is not observed in S CrA A, this is most probably due to the presence of strong veiling in the spectra of the source. 

The spectral classification of S CrA A and S CrA B is challenging.
Very few absorption lines are present in the spectrum of either component.
In addition strong veiling is present in the spectrum.
Our spectral classification is based on the absence of strong features that are characteristic of certain
spectral classes. 

First, given that broad Paschen {\it absorption} lines at 8000 \AA~ are not present in the spectra,
both components are not earlier than G.
Second, the absence of strong TiO molecular bands at 6185, 6720, and 7100 \AA, and CaH$_2$ (6385 \AA) 
and CaH$_3$ (6830 \AA~) features, indicates that primary and companion are earlier than M.   
In the case of S CrA A none of the strong absorption lines characteristic of a K star are observed.
The Ca I line at 6162 \AA~, 
the metallic lines between 6000 and 6600 \AA~, 
and the K I line at 7699 \AA~ are absent. 
We conclude that S CrA A (primary) is most likely a G-type star.
We assign a spectral range between G0 and G9, and adopt a G5Ve spectral type. 
In the case of S CrA B (companion) two features are observed, a weak K I line at 7699 \AA~ (EW$\approx$ 0.5 \AA),
three emission-filled Ca II absorption lines between 7000-8000 \AA~ (see in Fig. 4 the 8542 and 8662 Ca II lines,
where a narrow emission appears inside the core of the absorption line).
The simultaneous presence of the K I and Ca II features indicate that S CrA B is very likely a K-type star.
The presence of three {\it absorption} features at 8200 \AA~ and at the position of P(18) and P(17), 
that are only seen in K-type stars (comparison with STELIB template),
provide further evidence that S CrA B is of spectral type K.
Given that the strong Ca I line at 6162 \AA~ is absent in the spectra, S CrA B should be earlier than K7.
We conclude that S CrA B is in the spectral range K0 - K7, and adopt a K5Ve spectral type.
   
Prato et al. (2003) observed S CrA A\&B in the NIR.
They mention that S CrA A is a K3 type star and S CrA B a M0 type star (from Krautter 1991) and reported
the detection of CO bands at 2.3 $\mu$m.
The spectra of S CrA A\&B published by Prato et al. (Fig. 2),
show that the NIR spectrum is virtually flat in both components
and that none of the spectral features characteristic of G, K or M stars are present.
Prato et al. estimated that the veiling in the NIR was high ($r_K \approx$ 2-3)
\footnote{$r_K$ is the ratio of the magnitude of the K-band excess to the photospheric flux of the star at 2.2 $\mu$m.}.
We note that the optical spectrum provides better spectral
diagnostics for the classification than NIR spectra, as
in the optical range the photospheric contribution to the total
system flux is expected to be stronger than in the near-IR.
The discrepancy between Prato et al.'s spectral types and ours could be explained as 
the effect of the high NIR veiling.  

%
\begin{table*} 
\caption{Magnitudes of the companion stars in V, K, R and I bands. 
References: Baier et al. 1985 [B85], Herbst \& Shevchenko 1999 [HS99], Hillenbrand 1992 [HI92], Hipparcos catalogue [HIP], 
Kouwenhoven et al. 2005 [K05], Prato et al. 2003 [P03], Rositer 1955 [R55], Wycoff et al. 2006 [W06] 
}
\label{Mv}      
\centering                          
\begin{tabular}{lllllllll}  
\hline\hline                 
STAR&literature&&&&PSF spectrum&\\
&V[mag]&Ref&K[mag]&Ref&R[mag]&I[mag]&\\
\hline 
HD 144432 B&12.9$\pm$0.5     &R55     &9.0 $\pm$ 0.1 &$^a$&11.2 $\pm$ 0.2&10.1 $\pm$0.2\\
HD 150193 B&12.3$\pm$0.2     &HIP     &7.9 $\pm$ 0.1 &K05&10.9 $\pm$ 0.2&9.9 $\pm$0.2\\
KK Oph B   &13.0$\pm$0.3 $^b$&HS94    &8.1 $\pm$ 0.1 &HI92 $^c$&13.1 $\pm$ 0.2&11.7 $\pm$0.2& \\
S CrA A    &11.0$\pm$0.2     &B85,W06 &6.56 $\pm$ 0.06 &P03 & 14 $\pm$ 0.3 $^d$ & 13.0 $\pm$ 0.3 $^d$\\
S CrA B    &12.0$\pm$0.2     &B85,W06 &7.27 $\pm$ 0.08 &P03 & 14 $\pm$ 0.3 $^d$ & 13.2 $\pm$ 0.3 $^d$\\
\hline 
\end{tabular} 
\begin{flushleft} 
$^a$ From the acquisition IRTF image published in P\'erez et al. 2004. \\ 
$^b$ The combined brightness of KK Oph A\&B varies on short (days--months) and long (decades) timescales.
Herbig and Bell (1988) reported that in 1987 KK Oph B was 1 mag fainter than KK Oph A in the $V$ band.
To obtain the literature value of the $V$ magnitude of KK Oph B,
we used the photometry of KK Oph A in 1987 published in Herbst \& Shevchenko (1999).\\
$^c$ Pirzkal et al. (1997) reported that the companion to KK Oph A is 2.5 mag fainter in the K band. 
Here we used the Hillenbrand et al. (1992) magnitude measurement  ($K=5.64$ mag) of KK Oph A for estimating $K$ for the companion.\\
$^d$ The $R$ and $I$ magnitudes derived for S CrA A and B from the PSF are 1-2 magnitudes too faint to be consistent with the
$V$ and $K$ magnitudes reported in the literature and the spectral type of the sources. 
We include them here to illustrate the difference in flux of $\sim 0.2$ mag between S CrA A and B in the $R$ and $I$ bands.
\end{flushleft}
\end{table*} 
%
\subsection {Luminosity determination}
%
\begin{table*}
\caption{Details of the calculation of the luminosity of the companions.
The parameters $T_{eff}$ and the luminosity log $L$ of the primary are taken from the literature.
~$A_{\rm V}$ was deduced self-consistely from the $(V-R)$ and $(V-I)$ colors (see \S3.2). 
References: Acke et al. 2004 [A04], van Boekel et al. 2005 [B05], Hillenbrand et al. 1992 [H92], Leinert et al. 1997 [L97], Reipurth \& Zinnecker 1993 [RZ93].%
}
\label{HRDiagram}      
\centering                          
\begin{tabular}{llllllllllllllllllllllllllllll}  
\hline\hline                 
Star             &  &  SpT  &  log $T_{eff}$  & $d$ &  $A_{\rm V}$  & $V_{\rm Lit}-A_{\rm V}$ &  log $L$  & log $L_{\rm adopted}$ & Ref\\
              &   &      &log [K]&[pc]&[mag]& [mag]& [L$_{\odot}$]& [L$_{\odot}$]\\
\hline    
HD 144432     & A & A9IVe &3.87  &145 &0.17&                         &                  &  1.0 $\pm$ 0.1  & A04,B05\\
              & B & K4Ve  &3.66  &145 &1.4 $\pm$ 1.1 &11.5 $\pm$ 1.2 & -0.16 $\pm$ 0.5  & -0.2 $\pm$ 0.5  &\\
              &   & K7Ve  &3.61  &145 &0.9 $\pm$ 1.1 &12.0 $\pm$ 1.2 & -0.24 $\pm$ 0.5  &                 &\\
\\
HD 150193     & A & A2IVe &3.95  &150 &1.49 &                        &                  &  1.38 $\pm$ 0.1 & H92,A04,B05\\
              & B & F8Ve  &3.79  &150 &1.9 $\pm$ 1.1 &10.4 $\pm$ 1.1 &  0.15 $\pm$ 0.4  &  0.13 $\pm$ 0.4 & \\
              &   & G5Ve  &3.76  &150 &1.8 $\pm$ 1.1 &10.5 $\pm$ 1.1 &  0.11 $\pm$ 0.4  &                 &\\
\\
KK Oph        & A & A6Ve  &3.92  &160 &1.6           &               &                  &  1.3 $\pm$ 0.1  & H92,L97 \\
              & B & G5Ve  &3.76  &160 &2.8 $\pm$ 1.1 &10.2 $\pm$ 1.1 &  0.32 $\pm$ 0.4  &  0.3 $\pm$ 0.4  &   \\
              &   & G8Ve  &3.74  &160 &2.8 $\pm$ 1.1 &10.2 $\pm$ 1.1 &  0.30 $\pm$ 0.4  &                 &   \\
\\
S CrA         & A & G0Ve  &3.78  &140 &3.2 $\pm$ 1.6 &7.8  $\pm$ 1.6 &  1.14 $\pm$ 0.6  &  1.1 $\pm$ 0.6  & RZ93\\
              &   & G9Ve  &3.73  &140 &3.0 $\pm$ 1.6 &8.0  $\pm$ 1.6 &  1.08 $\pm$ 0.6  &                 & \\
              & B & K0Ve  &3.72  &140 &2.9 $\pm$ 1.6 &9.1  $\pm$ 1.6 &  0.67 $\pm$ 0.6  &  0.6 $\pm$ 0.6  & \\
              &   & K7Ve  &3.61  &140 &1.9 $\pm$ 1.6 &10.1 $\pm$ 1.6 &  0.53 $\pm$ 0.6  &                 & \\
\hline  
\end{tabular} 
\vfill
\end{table*}
%
For calculating the luminosity of the companions,
their distance, 
visual magnitude, 
extinction and 
bolometric correction
are required.
Given that all the companions are emission line stars
it is unlikely that they are field background objects. 
Since the companions are close (in projected distance) to the primaries,
we assume that they are at the same distance. 

The visual magnitudes $V$ are available from the literature (see Table 3).
To determine the visual extinction $A_{\rm V}$,
we calculated the $R$ and $I$ magnitudes from the total number of photons received in the spectra,
and derived $A_{\rm V}$ self-consistently using the $(V-R)_i$ and $(V-I)_i$ colours intrinsic to 
the spectral-type(s) assuming a standard galactic extinction law: 
$0.37A_{\rm V} = (R-I)_{obs}+(V-R)_i-(V-I)_i$ (Fluks et al. 1994).

For determining $R$ and $I$, in the cases of HD 144432 B, HD 150193 B and KK Oph B,
we used the following procedure: 
First, in a flat-fielded, background emission and cosmic-ray corrected science target exposure, 
the total number of photons inside the PSF in the wavelength band was counted
(for $R$ band we used the 1200R grism exposure and for $I$ band the 1028z grism).
The flux of the target $F_{TARGET}$ was obtained normalizing the total number of counts by the frame's exposure time.
Second, the flux of the standard star $F_{STD}$ was derived applying the same procedure to a standard-star frame.
The $R$ or $I$  magnitudes were estimated using the 
classical flux/magnitude relation:  
$M_{TARGET} = M_{STD}~-~2.5~\times~log(F_{TARGET}/F_{STD})$.
We employed a PSF window 30 pixels-wide and used the star HD 156026 
(V=6.3 mag, R=5.4 mag, I=4.8 mag, seeing at the time of observation 0.6")
as photometric standard.  
The errors on the derivation of $R$ and $I$ are maximum 0.2 mag:  
0.1 mag by photon noise and systematics and maximum 0.1 mag due to slit differences 
in the seeing between the target and standard star observations.
In the three objects extinction towards the companion is larger than towards the primary.

The method used for the extraction of the spectra of S CrA A and S CrA B did not conserve
the fidelity of the fluxes (cf \S 2.2). 
Therefore, a different approach was required for determining $F_{TARGET}$ on these sources.
The two-dimensional science frame was converted to one-dimension by summing  
the counts in the dispersion direction.
Two Gaussians with the same FWHM were fitted to the double-peaked profile obtained. 
The relative area of each Gaussian normalized by the exposure time determined the flux of each of the binary components.
Using the flux/magnitude relation, 
the $R$ and $I$ magnitudes were calculated for each component.
The $R$ and $I$ obtained are up to 2 mag too faint to be compatible with previous measurements of $V$ and the spectral
type of the sources.
That may be due to the intrinsic variability of S CrA (A\&B) ($V$=10.5 - 12.4 mag; de Winter et al. 2001).
Although not adequate for absolute photometry, the estimates of $R$ and $I$ in S CrA A and S CrA B do 
allow relative photometry, and permit us to obtain an estimate of the visual extinction towards the sources.
We find that the companion is $\approx 0.2\pm0.3$ magnitudes fainter than the primary in the $R$ and $I$ band.
 
The absolute magnitude $M_{\rm V}$ of the companions was derived from our estimation of $A_{\rm V}$ and
$V$-band measurement from the literature.
Given that the extinction-corrected visual magnitude estimate and the bolometric correction
depend on the spectral type, we calculated the luminosity for the two extremes of the spectral range derived 
for each companion.
The spectral type $(V-R)_i$ and $(V-I)_i$ colors, the bolometric correction and the effective temperature calibrations were taken from Table A5 of Kenyon \& Hartmann (1995).
The error in the luminosity is determined by the error in the $M_V$ magnitude estimation and the range of the bolometric correction.
The typical error in the determination of ${\rm log}(L/L_{\odot})$ was 0.4 to 0.6.  
A summary of the calculation of the companions' luminosities is given in Table 4. 

\begin{figure}
\centering
\includegraphics[angle=0,width=0.5\textwidth]{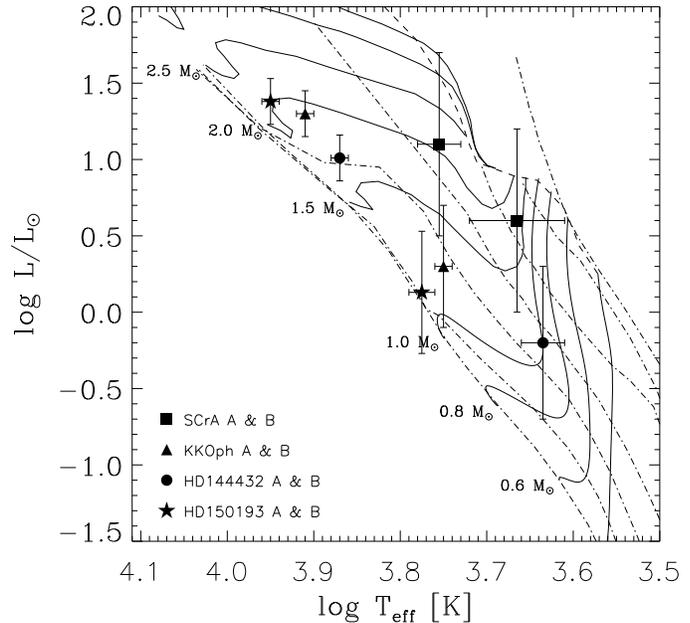}
      \caption{
HR diagram of the primaries and the companions.
In each pair, the upper star is the primary and the lower one is the companion.
The theoretical evolutionary tracks are from Palla and Stahler (1993).
The primaries and companions are located on similar isochrones within the errors.
The isochrones correspond to
0.1, 0.3, 1, 3, 10, 30 and 100 Myr respectively.}
         \label{HR diagrams}
\end{figure}
\subsection{Primaries and companions in the HR diagram}
With the estimates of the effective temperature $T_{eff}$ and the luminosity $L$, 
we localized the primaries and their companions in the HR diagram (see Figure 5). 
To calculate the mass and the age of the systems,
we used the theoretical evolutionary tracks of Palla \& Stahler (1993).
The parameters of the systems (masses and ages) are summarized in Table 5.
The first column shows the mass of each source; 
the second, the age range for each star.
Primaries and companions are within the errors coeval objects.
Assuming coevality, the system age (third column) is
is the age interval that is {\it simultaneously} consistent 
with the age derived for the primary and for the companion. 
It is given by the 1-sigma confidence level interval of the age probability distribution 
obtained by multiplying the age probability distribution of the primary and companion 
(assuming symmetric Gaussians in the log(age) space).

Fukagawa et al. (2003) using photometry and the evolutionary tracks
of Siess et al. (2000) derived a mass of 1.4 M$_\odot$ and age of 5 Myr for HD 150193 B.
We demonstrated in \S3.1.2 that HD 150193 B is a F8-G5 star and not a mid-K type star. 
Taking into account the uncertainly in the luminosity estimate, 
our spectral-type range rules out ages smaller than 10 Myr for HD 150193 B.  

\section{Discussion}
We obtained spatially resolved optical spectroscopy of the close 
companions to  the Herbig Ae/Be stars HD 144432, HD 150193, KK Oph and the 
T Tauri star S CrA A.
Our first objective was to find out whether the visual companions 
are physically associated objects or background sources.
For answering this, 
we determined their spectral types.
Using $V$ photometry from the literature and estimated $A_{\rm V}$ 
values from our data, 
we determined their luminosities and their location in the HR diagram. 
The results of our observations are:\\
i) The spectra of all the companions are characterized by H$_\alpha$ 
in emission and in some cases also strong Ca II in emission.\\
ii) The spectra of the companions are characterized by Li I in absorption\\
iii) Several companions have early spectral types.\\
iv) Assuming the same distance, primaries and companions are in the pre-main-sequence zone of the HR diagram
and -- within the errors -- they are localized on similar isochrones as their primaries.\\  

Let us now address the question of whether the companions are physically associated with the primary.
First, it is unlikely that the companions are background sources. 
Background sources usually do not have emission lines and in general are not early-type stars.
Second, the companions are very likely young.  
HD 144432 B and KK Oph B are pre-main-sequence stars with masses $<$ 1.2 M$_{\odot}$ and show Li I in absorption.
Stars with masses smaller than 1.2 M$_{\odot}$ are expected to have photospheres significantly depleted in
lithium by the time they reach the main-sequence (Stahler \& Palla  2004).
Therefore, the presence of Li I in HD 144432 B, KK Oph B combined with their location above the ZAMS indicate that they are young. 
In addition, KK Oph is an actively accreting T Tauri star (H$_\alpha$ EW$\approx$ $<<$ -5 \AA). 
HD 150193 B has about 1.2 M$_{\odot}$ and is located low in the HR diagram close to the ZAMS. 
In consequence, the detection of Li I is a less obvious sign of youth.
However, HD 150193 B is an early-type star (F9V) 
with accretion signatures (H$_\alpha$ and Ca II in emission)
and X-ray emission (see below).
Therefore, HD 150193 B seems still to be young.
S CrA B (M $>$ 1.2 M$_{\odot}$) must be young because its position in the HR-diagram and its strong accretion signatures.  

These results combined with the small projected separation from their primaries,
{\it provide strong evidence that
HD 144432 B, HD 150193 B and KK Oph B are physically 
associated T Tauri stars to Herbig Ae/Be primaries, 
and confirm that S CrA is a T Tauri binary}.
KK Oph B and S CrA B are actively accreting T Tauri stars and are very likely surrounded by disks.
HD 150193 B is an early-type T Tauri star with a low accretion rate,
HD 144432 B is a weak-line T Tauri star without strong evidence for a disk.  
Follow up studies analyzing the proper motion of the objects,
as reported in the case of HD 144432 A \& B (P\'{e}rez et al. 2004),
and HD 150193 A\&B (Fukagawa et al. 2003) 
will be extremely useful to confirm that the systems are true binaries, 
and to derive their orbital parameters.

\begin{table}
\caption{Masses and ages derived for the companions and primaries from the HR diagram.
The system age is the age interval that is simultaneously consistent with the age derived for the 
primary and the companion (see \S 3.3).}             
\label{table:5}      
\centering                          
\begin{tabular}{l l l r c}        
\hline\hline                 
Star      &   & Mass       & Age   & System age\\    
          &   & [M$_{\odot}$] & [Myr] & [Myr]     \\
\hline
\\[0.5mm]                        
HD 144432 & A &  1.6 $\pm$ 0.1 &   9 $^{+7}_{-2}$   & 8$^{+3}_{-1}$\\[1mm]      
          & B &  1.0 $\pm$ 0.2 &   4 $^{+26}_{-3}$  & \\[2mm]
HD 150193 & A &  2.0 $\pm$ 0.1 &   7 $^{+2}_{-1}$   & 10$^{+1}_{-1}$\\[1mm]
          & B &  1.2 $\pm$ 0.2 &  30 $^{+1000}_{-20}$ & \\[2mm]      
KK Oph    & A &  1.6 $\pm$ 0.1 &   7 $^{+2}_{-1}$   & 7$^{+1}_{-0.5}$\\[1mm]
          & B &  1.1 $\pm$ 0.2 &  12 $^{+18}_{-7}$  & \\[2mm]
S CrA     & A &  2.2 $\pm$ 0.7 &   2 $^{+8}_{-1}$   & 2$^{+0.5}_{-1}$\\[1mm]
          & B &  1.3 $\pm$ 0.7 &   1 $^{+3}_{-0.9}$ & \\[2mm]
\hline                                   
\end{tabular}
\end{table} 

The presence of T Tauri companions physically associated with Herbig Ae/Be stars 
raises interesting questions concerning the potential of such objects to form planets
and, in a more general sense, about the formation scenarios of the Herbig Ae/Be stars themselves.

Growing observational evidence for binary and multiple systems
points towards ``isolated'' Herbig Ae/Be stars being the exception rather than the rule
(e.g., Chen et al. 2006; Baines et al. 2006).

Let us now consider the X-ray emission.
In our sample of Herbig stars, X-ray emission has been observed towards HD 150193 by ROSAT and Chandra
\footnote{KK Oph and  HD 144432 have no reported X-ray detections.}.
Feigelson et al. (2003), Skinner et al. (2004) and Stelzer et al. (2006) 
imaged HD 150193 with Chandra and detected two X-ray sources:
a faint X-ray northern source and a bright X-ray southern source.
The nearly identical separation and position angle with respect to the IR image from Pirzkal et al. (1997),
suggest a one-to-one correspondence between the two sources seen in the X-ray and IR images. 
Our observations show that the IR companion is a T Tauri star.
This result provides further evidence that the X-ray bright source is HD 150193 B.
In the case of the T Tauri star S CrA A\&B, 
X-rays have been detected by ROSAT, XMM and Chandra.
Chandra deep (160 ksec) data of the region show that the source emission appears to be only due
to the primary (Forbrich \& Preibisch 2007, submitted to A\&A).

Let us now move to the structure of protoplanetary disks.
There are several potential effects of a companion.
The Herbig Ae/Be stars HD 100453, HD 34700, HD 169142, MWC 1080, HD 35187, HD 141569 have no 10  $\mu$m silicate feature,
and they all have companions.
Chen et al. (2006) suggested that the lack of 10 $\mu$m silicate may be linked to the presence of 
close companions. 
Although further investigations are needed to establish whether Herbig stars that lack 
the 10 $\mu$m silicate feature show a higher binary frequency than stars with silicate feature,
our observations do not support this as a general conclusion.
All Herbig stars in our sample (HD 144432, HD 150193, KK Oph),
do show silicate emission (Acke \& van den Ancker 2004),
yet they also show close companions 
(HD 144432 B, separation $\approx 200$AU, 
HD 150193 B, separation $\approx 160$AU,
KK Oph separation $\approx 240 $AU).
Some of our sources (HD 150193) are very similar in other properties
(age, mass, distance to the companion)
to sources in the Chen et al. sample (i.e HD 100453, separation $\approx 120$ AU) .
So we can find clear examples of sources with close companions which show evidence for 
the presence of relatively small dust grains in the surface layers of their protoplanetary disks, 
as well as sources with close companions in which these small dust grains are absent, either
due to settling or grain coagulation.  
Other parameters than binarity must thus play a role
in the processing of dust in the disks around Herbig Ae/Be stars.
   
Lets us now consider the influence of binarity on the disk's lifetime.
Theoretically (e.g. Artymowicz and Lubow, 1994) and observationally (e.g. Bouwman et al., 2006), 
it has been suggested that in low-mass close (projected separation $< 20$ AU) T Tauri binaries,
the disks have shorter lifetimes compared to single-star systems. 
Binarity may well be a key issue in Herbig stars too. 
In our sample (projected separation $\approx$~200~AU), 
three of the four systems studied (HD 144432, HD 150193, KK Oph) have ages $>$ 7 Myr,
suggesting that they may have retained their disks for a much longer time than typical for a young star.
Our sample is small, it does not represent a complete sample, 
and we should be cautious about extrapolating our findings. 
However, 
it could be that Herbig stars may retain their disks much longer than other young objects 
because of a companion's influence. 
Further research is required to determine if Herbig stars with close companions
consistently retain their disks longer than isolated objects.

Another interesting point is the influence of the Herbig primary on the companion's disk
lifetime.
All our systems have a mass ratio of $\approx$0.6 and a similar separation,
S CrA A\&B is 2 Myr old and both stars show evidence for disks.
KK Oph A\&B is 7 Myr old and the companion is still an active accreting object.
HD 144432 A\&B is 8 Myr old and the companion is apparently disk free.
HD 150193 A\&B is 10 Myr old and the companion is a weak accretor.
In our sample the companion's disk has dissipated after 8 Myr. 
As in the case of the Herbig primaries, our companions retained their disks longer than
typical for a young star. Now, HD 150193 B at 10 Myr still has its disk and HD 144432 B (8 Myr) not. 
One important difference between HD 150193 B and HD 144432 B is that the former is more massive.
We speculate that at similar primary/secondary ratio and separation ($\sim$~200 AU), 
the secondary may be more likely to retain its disk longer the higher its mass is. 
Studies with larger samples with a variety of mass ratios and separations are required to 
statistically test these trends.  

%
\begin{acknowledgements}     
A.C. is grateful to J.Hernandez for providing the calibration curves
for spectral classification, 
and for discussions concerning the spectral classification of pre-main-sequence stars.
A.C. would like to thank K.T. Hole and C. Grady for useful comments on the manuscript,
and to J. Bouwman for helpful discussions.
We are very grateful to the referee for comments and suggestions that helped to improve the paper.
This research has made use of the SIMBAD database
operated at CDS, Strasbourg, France. 
\end{acknowledgements}
%
%

%
%
\end{document}